\documentclass[aps,prd, groupedaddress,amsmath,amssymb]{revtex4}

\usepackage{epsfig}
\usepackage{dcolumn}
\usepackage{bm}
\usepackage{slashed}

\begin{document}

\def\bea{\begin{eqnarray}}
\def\eea{\end{eqnarray}}
\def\as{\alpha_s}
\def\zp{Z^\prime}
\def\zpmd{\tilde{Z}^\prime_\mu}
\def\zpmnd{\tilde{Z}^\prime_{\mu\nu}}
\def\zpmu{\tilde{Z}^{\prime \mu}}
\def\zpmnu{\tilde{Z}^{\prime\mu\nu}}
\def\nn{\nonumber}
\def\eps{\varepsilon}
\def\d{{\rm d}}

\preprint{FREIBURG PHENO-08-09}
\title{High-dimensional $Z^\prime$ phenomenology at hadron colliders}
\author{Benjamin Fuks}
\author{Jochum J. van der Bij}
\email[]{jochum@physik.uni-freiburg.de}
\author{Qingjun Xu}
\affiliation{Physikalisches Institut, Albert-Ludwigs-Universit\"at
 Freiburg, Hermann-Herder-Stra\ss{}e 3, D-79104 Freiburg i.Br.,
 Germany}
\date{\today}
\begin{abstract}
We study the phenomenology of a $Z^\prime$-boson field coupled
to hypercharge. The $Z^\prime$ propagator has a nontrivial
K\"all\'en-Lehmann spectral density due to the mixing with
a higher-dimensional inert vector field. As a consequence
detection possibilities at hadron colliders are reduced.
We determine the range of parameters where this field can be
studied at the Tevatron and the LHC through its production cross section via the
Drell-Yan mechanism.

\end{abstract}
\pacs{11.10.Kk, 12.60.Cn, 13.85.Ni}

\maketitle


\vspace*{-60mm}
\noindent FREIBURG PHENO-08-09\\
\vspace*{50mm}

\section{Introduction}
\label{sec:1}

The present day data from  high-energy colliders like LEP and 
Tevatron show that almost all measurements are described by the standard model
at the loop level. Therefore  extensions of the standard model 
tend to be strongly constrained. Typically such extensions will spoil
the agreement with the data through a variety of effects, one of the
most important of which is the appearance of flavor changing neutral
currents. Even the most popular extension, the minimal supersymmetric
extension of the standard model, has to finely tune a  number of
parameters. This leaves only one type of extensions that are safe,
namely, extensions with singlet particles, since singlets do not affect
precision variables through loop effects. Since the discovery that
neutrinos are massive, it is clear that singlet fermions play a role
in nature. Moreover arguments from cosmology suggest that singlet
scalars could be important 
\cite{McDonald:1993ex,Bento:2001yk,Burgess:2000yq,Davoudiasl:2004be,vanderBij:2006ne}. %
In a recent analysis of the Higgs-search
data from LEP2 it has been pointed out that such singlet scalars may
already have been seen as a smeared-out Higgs boson
\cite{vanderBij:2006pg}. Given this situation it is natural to ask
whether also singlet vector bosons can be present.\\

Therefore, we decide to study  renormalizable extensions of the
standard model containing extra gauge bosons, but no extra fermions or
scalars. Demanding the absence of anomalies in the gauge currents, additional vector
bosons can couple to linear combinations of hypercharge ($Y$) and
the difference of baryon and lepton quantum numbers (B\textrm{-}L). However, 
they can have an arbitrary mixing with the standard model hypercharge field. 

In this paper we limit ourselves to the case where the vector bosons couple
to the hypercharge only. Such models are, after a redefinition of the gauge fields,
equivalent to the standard model with a number of inert gauge bosons, that only
couple to the standard model via the mixing with the hypercharge field. 
Since the extra vector bosons are singlets, their mass does not necessarily
have to arise from spontaneous symmetry breaking, but can be put in by hand without
violating renormalizability. For phenomenological purposes the differences
are minor. Both cases have been studied in the literature
\cite{Appelquist:2002mw,Kors:2005uz,Chang:2006fp,Feldman:2006wb,Feldman:2006ce,
Ferroglia:2006mj,Coriano:2008wf,Chanowitz:2008ix}.
However if the mass is not generated by the Higgs mechanism, an interesting
possibility arises when an infinity of fields is present. 
It is possible that the extra vector bosons move in more than 
four dimensions \cite{Ferroglia:2006mj}.
This will not violate the principle of renormalizability, if the number of
dimensions
is smaller than or equal to six; also fractional dimensions are possible.
As a consequence one finds, after diagonalizing
the mass matrix,  a physical neutral $Z^\prime$-boson field coupled to hypercharge. 
However, this field does not have a single particle peak but has a well-defined
nontrivial continuous K\"all\'en-Lehmann spectral density. Therefore,
the search for such a field
at hadron colliders becomes more complicated, since the signal is smeared
 out over a large invariant-mass range and no clear peak is present.
Because of their higher-dimensional nature we call this class of models Heidi models
(high-D for higher dimension).\\

It is the purpose of this paper to determine the range of parameters
of the model that can be studied at hadron colliders.
We consider a minimal version of Heidi models with a reduced set of free
parameters, namely, an extension of the standard model with one single additional
inert gauge boson mixing with the standard model hypercharge
field, which affects the usual standard model neutral gauge bosons. While the photon is
protected from further mixing by the exact electromagnetic symmetry, no
conserved quantum number prevents the $Z$ boson to mix with the additional $Z^\prime$
boson.
Theoretical calculations of various electroweak observables are then modified by
the presence of the additional field, which may change the fits of the
electroweak precision data and affect the bounds on the Higgs mass
\cite{Chanowitz:2001bv, Chanowitz:2002cd, Ferroglia:2006mj, Chanowitz:2008ix}.
 According to this point of view, measurements 
from LEP1 and SLD at the $Z$ pole \cite{lepewwg} were analyzed in
Ref.\ \cite{Ferroglia:2006mj}, both for a four-dimensional and
for the Heidi $Z^\prime$ models. The analysis on the technical level is not significantly
different for the four-dimensional and the higher-dimensional case.
The results of this analysis were recently confirmed  in Ref.\ \cite{Chanowitz:2008ix}.
Fitting the data leads to constraints on the allowed parameter range of the Heidi models.
We will restrict ourselves to the limits given in Ref.\ \cite{Ferroglia:2006mj},
which are on the conservative side.\\

The Drell-Yan-like production of a lepton pair at hadron colliders plays an
important role in the present and future experimental program at the Fermilab
Tevatron and CERN LHC colliders thanks to its large cross section and clean
signature of the final state. This is also the primary discovery mode for
additional neutral gauge bosons and has already been widely studied in
the literature for the Tevatron and the LHC
\cite{Dittmar:2003ir,Carena:2004xs,Fuks:2007gk,Leike:1998wr,Langacker:2008yv}. It has also
been shown that the forward-backward asymmetry due to the interference between
the three  different exchange modes ($\gamma$, $Z$, and $Z^\prime$) could be
an interesting observable
\cite{Collins:1977iv,Langacker:1984dc,Rosner:1995ft,Dittmar:1996my}.
Therefore, in this
work, we investigate for the first time the possible effects of a Heidi 
$Z^\prime$ on the Drell-Yan dilepton invariant-mass spectrum and
the forward-backward asymmetry, both at the Tevatron and the
LHC.

Because of the important theoretical uncertainties coming from QCD effects and
missing large higher-order contributions, leading-order results for the Drell-Yan
process, with or without additional $Z^\prime$, are quite unreliable.
A systematic approach to this problem is based on perturbation theory truncated
at the next-to-leading order (NLO) \cite{Altarelli:1979ub,Aurenche:1980tp} or
next-to-next-to-leading order \cite{Hamberg:1990np} in the strong coupling
constant $\as$. In kinematical regions where higher-order contributions
are enhanced due to the soft and collinear parton emission, soft-gluon
resummation to all orders in $\as$ can be performed and matched with the fixed
order predictions in order to improve the description of
observables under consideration. However, a general  procedure
 has not yet been
developed and the resummation has to be done explicitly process by process. An
alternative way of treating the soft-gluon emission is the parton shower,
which approximates the full resummation calculation. However,
the parton shower is   universal
for all  processes and can be implemented easily in the
analysis chains of the Tevatron and LHC experiments. Consequently, we implement
in this work the Heidi $Z^\prime$ bosons in the Monte Carlo generator MC@NLO
\cite{Frixione:2002ik}, generalizing the modified MC@NLO program including
$Z^\prime$ gauge bosons inspired by grand unified theories of Ref.\
\cite{Fuks:2007gk}, which allows 
us to match NLO perturbative calculation with the parton shower of the Monte
Carlo generator HERWIG \cite{Corcella:2000bw} including color coherence effects
and azimutal correlations within and between jets. We subsequently compare 
the Monte Carlo
predictions to the experimental results from the Tevatron
for 
the invariant-mass spectrum and
forward-backward asymmetry of the Drell-Yan production of lepton-pairs.
 In addition we study the sensitivity of the LHC
experiments to the Heidi models for this channel.\\

The remaining part of the paper is organized as follows. In Sec.\
\ref{sec:2}  we first describe the  theoretical framework, i.e.\ the
minimal Heidi  
$Z^\prime$ model, and discuss the implementation of $Z^\prime$-boson production in 
the MC@NLO generator. Section \ref{sec:3} is devoted to the numerical
analysis of the experimentally allowed Heidi parameter space and the
investigation of the sensitivity of the LHC experiments to the model. Our
conclusions are given in Sec.\ \ref{sec:4}.

\section{Theoretical framework}
\label{sec:2}


\subsection{The minimal Heidi model}

There is a large class of models containing extra neutral vector bosons
\cite{Leike:1998wr,Langacker:2008yv}. The simplest extension consists of the
 standard model
plus one additional $U^\prime(1)$ symmetry, the extended gauge group being  $SU(3)_C
\times SU(2)_L \times U(1)_Y \times U^\prime(1)$. 

We associate with the extra symmetry a $\zpmd$ field that is a singlet under the standard model
gauge group. Moreover, we assume that no particle carries a $U^\prime(1)$ charge.
Therefore, this $\zpmd$ field is inert and its presence can only be noticed, because it
can mix with the Abelian field $\tilde{B}_{\mu}$ associated with the $U(1)_Y$ symmetry. 
Because it does not couple to any fermions or Higgs fields, the additional
$\zpmd$  field is allowed to move in $D$ dimensions. As long as  
$D \leq 6$ the theory will stay renormalizable. In contrast, the $\tilde{B}_{\mu}$ field
must be four-dimensional, because of its couplings
to the fermionic fields. Therefore, in this basis, the hypercharge basis, we have two
fields, a four-dimensional field coupled to the hypercharge and a higher-dimensional
field coupled to nothing. However, the two fields can be mixed by a mass term. In order
to describe the physics correctly, one has to transform to the mass basis.
Allowing the $U(1)$ fields to have masses and mass-mixing terms, one finds
for the gauge field part of the Lagrangian in the hypercharge basis 
\cite{Ferroglia:2006mj}:

\bea
 -\mathcal{L}_{\rm gauge} =  \frac{1}{4} \tilde{B}_{\mu\nu} \tilde{B}^{\mu\nu} 
                           + \frac{1}{2} m_4^2 \tilde{B}_{\mu} \tilde{B}^{\mu} 
			   + \frac{1}{4} \zpmnd \zpmnu 
			   + \frac{1}{2} m_D^2 \zpmd \zpmu
			   + M^{4-\frac{D}{2}}_{\rm mix} \zpmd \tilde{B}^{\mu}.~
\eea
We have a four-dimensional mass $m_4$, a higher-dimensional mass $m_D$,
and a scale $M_{{\rm mix}}$ that connects the four-dimensional fields with
the higher-dimensional ones.
The two massive hypercharge fields $\zpmd$ and $\tilde{B}_{\mu}$ are transformed to the
mass-eigenstates $\zp_\mu$ and $B_{\mu}$, the latter being massless since the
electromagnetic symmetry is exact.
As a consequence, the three parameters $m_4$, $m_D$, and $M_{\rm mix}$ are not
independent and must satisfy the condition 
\renewcommand{\arraystretch}{1.5}
\bea 
  \left\{ \begin{array}{l c l l}
    m^2_4\, m_D^{6-D} = \mu_{\rm lhd}^{8-D}\, & {\rm~for~}D\neq 6\\
    m^2_4 + \mu_{\rm lhd}^2 \ln\frac{m_D^2}{\mu_{\rm lhd}^2} = 0\, &
      {\rm~for~}D=6
  \end{array}~,~ \right. \label{eq:condphot}
\eea
where we have introduced the quantity $\mu_{\rm lhd}$ in reference to the mixing of
the low-dimensional and high-dimensional fields 
\bea
 \mu_{\rm lhd}^{8-D} \equiv 
   \frac{\Gamma\big[3-\frac{D}{2}\big]}{(4 \pi)^{\frac{D}{2}-2}}\,
   M_{\rm mix}^{8-D} ~.~ 
\eea

By compactifying the higher dimensions and subsequently taking the continuum
limit, one can derive the hypercharge-boson propagator which is of the form 
\renewcommand{\arraystretch}{1.6}
\bea
 D^{\tilde{B}\tilde{B}}_{\mu\nu}(q^2) = -i g_{\mu\nu} 
   D^{\tilde{B}\tilde{B}}(q^2) &=& \left\{ \begin{array}{l l} 
     -i g_{\mu\nu} \bigg[ q^2 - m_4^2 + \mu_{\rm lhd}^{8-D} \big(q^2-m_D^2
     \big)^{\frac{D-6}{2}} \bigg]^{-1} \, & {\rm~for~} D \neq 6 \\ 
   -i g_{\mu\nu} \bigg[ q^2 - m_4^2 - \mu_{\rm lhd}^2 \ln
     \frac{m_D^2-q^2}{\mu_{\rm lhd}^2} \bigg]^{-1} \, & {\rm~for~} D = 6
   \end{array}\right. ~,~ \label{eq:prop}
\eea and its nontrivial corresponding K\"all\'en-Lehmann spectral density
\bea
\rho(s) = -\frac{1}{2 \pi i} \lim_{\eps\to0} \Big[
   D^{\tilde{B}\tilde{B}}(s + i\eps) -
   D^{\tilde{B}\tilde{B}}(s - i\eps)  \Big] ~,~
\eea 
which becomes in the four-, five- and six-dimensional cases
\bea
 \rho_4(s) 
 &=& \frac{m_D^2}{m_D^2 + m_4^2}\, \delta\big(s\big) +   
   \frac{m_4^2}{m_D^2 + m_4^2} \delta\big(s-m_D^2 - m_4^2\big)~,~\nn\\
 \rho_5(s) &=& \frac{2 m_D^2}{2 m_D^2 + m_4^2}\, \delta\big(s\big) +
   \frac{\theta\big(s-m_D^2\big)}{\pi}\, \frac{m_4^2\, m_D\, 
   \big(s-m_D^2\big)^{\frac{1}{2}}}{\big(s-m_4^2\big)^2\, \big(s-m_D^2\big) +
   m_4^4\, m_D^2}~,~\nn\\
 \rho_6(s)&=& \frac{m_D^2}{m_D^2 + \mu_{\rm lhd}^2} \delta\big(s\big) +
   \theta\big(s-m_D^2\big)\, \frac{\mu_{\rm lhd}^2}{\Big(s-\mu_{\rm lhd}^2
   \ln\frac{s-m_D^2}{m_D^2}\Big)^2 + \mu_{\rm lhd}^4\, \pi^2} ~,~
\eea respectively. 
For the three choices of the number of dimensions $D$, we recover models with
one massless excitation which becomes the photon after the breaking of the
electroweak symmetry. This is guaranteed by the condition of Eq.\
(\ref{eq:condphot}). However, while in the four-dimensional case we have a 
second peak corresponding to a massive resonance located at
\bea
 m_{\zp}^2 \equiv m_D^2 + m_4^2 ~,~ \label{eq:zpmass}
\eea for five- and six-dimensional singlet fields, we now have  a massive
continuum. As stated above, the presence of this new field affects the standard model
$Z$-boson through a mixing which can be described by a quantity $a_Y$
defined by \cite{Ferroglia:2006mj} 
\bea
 a_{Y,4} &=&  \sin^2\theta_W \frac{m_Z^2\, m_4^2}{m_D^2\,(m_4^2+m_D^2)} ~,~\nn\\ 
 a_{Y,5} &=& \sin^2\theta_W \int_{m_D^2}^\infty \frac{m_Z^2\, \d s}{2 \pi\, m_D\, s}\,
   \frac{\big(2 m_D^2+m_4^2\big)\, m_4^2\, \big(s-m_D^2\big)^{\frac{1}{2}}}
   {\big(s-m_D^2\big)\,\big(s-m_4^2\big)^2 + m_D^2\, m_4^4 } ~,~\nn\\
 a_{Y,6} &=&  \sin^2\theta_W \int^{\infty}_{m_D^2} m_Z^2\, \d s \frac{m_D^2 + \mu_{\rm
 lhd}^2}{m_D^2 \,s}\, \frac{\mu_{\rm lhd}^2}{\Big[s - \mu_{\rm lhd}^2 \ln
 \frac{s - m_D^2}{m_D^2}\Big]^2 + \pi^2 \mu_{\rm lhd}^4}~,~
\eea for the four-, five-, and six-dimensional cases,
respectively. This quantity enters directly the vector and axial-vector coupling
strengths of the $Z$ boson to fermions, which read now
\bea
  v_f &=& e_f \big(\sin^2\theta_W - a_Y\big) - \frac{1}{2} T^3_f \big(1 - a_Y \big)~,~\nn\\
  a_f &=& - \frac{1}{2} T^3_f \big(1 - a_Y \big) ~.~
\eea $T_f^3$ and $e_f$ denote the weak isospin quantum number and the
electric charge of the fermion $f$, and $\sin^2\theta_W$ is the squared sine of the
electroweak mixing angle. The coupling strengths of
the new $Z^\prime$ state to fermion are simply given by the hypercharge 
\bea
  v_f^\prime &=& F(m_4, m_D) \, \sin\theta_W \big(\frac{T^3_f}{2} - e_f\big)~,~ \nn\\
  a_f^\prime &=& F(m_4, m_D) \, \sin\theta_W \frac{T^3_f~}{2},~ \label{eq:zpc}
\eea with a prefactor $F$ depending on the four-dimensional and high-dimensional
masses.

One can interpret the propagator of Eq.(4) in two ways. The simplest way to treat the
theory is to consider the field as a hypercharge gauge field with a nontrivial
K\"all\'en-Lehmann spectral density. The other interpretation is to say 
that one has many (in this case a continuum of) single hypercharge coupled
$Z^\prime$
fields, however, with a reduced coupling constant. The latter interpretation
would be natural if we had an integer number of compact higher dimensions. 


\subsection{Heidi $Z^\prime$production at hadron colliders}

The Drell-Yan-like production of a lepton pair plays an
important role in the present and future experimental program of hadron
colliders thanks to its large cross section and clean signature of the final
state. Since any observables related to this process are highly sensitive to the
existence of an additional neutral gauge boson, this is their primary discovery
mode. Following the conventions for the Drell-Yan process of Ref.\
\cite{Aurenche:1980tp}, which are the ones used in the Monte Carlo generator
MC@NLO, the corresponding squared matrix element reads

\bea
   q\bar{q} ~ {\rm or} ~ qg \to &\gamma,Z,Z^\prime + X & \to l^-l^+ + X ,~ 
\eea and can be written as
\bea
 \overline{|{\cal M}_i|^2}(q\bar{q}~{\rm or}~qg\to&\gamma,Z,Z'&\to l^-l^++X)
 ~=~ {1\over4}\,e^4\,C_i\,\bigg\{ {e_q^2\over M^4}T_i|^{1,0}_{1,0}\nn\\
 &+& {1\over\sin^4\theta_W\cos^4\theta_W} 
     {1\over(M^2-m_Z^2)^2+(\Gamma_Zm_Z)^2} T_i|^{A_l,B_l}_{A_q,B_q} \nn\\
 &+& {1\over\sin^4\theta_W\cos^4\theta_W} 
     \Big|D^{\tilde{Z}^\prime\tilde{Z}^\prime}(M^2)\Big|^2\,
      T_i|^{A'_l,B'_l}_{A'_q,B'_q} \nn\\
 &-& {2e_q\over M^2} {1\over\sin^2\theta_W\cos^2\theta_W} 
     {M^2-m_Z   ^2\over(M^2-m_Z   ^2)^2+(\Gamma_Z   m_Z   )^2} 
     T_i|^{v_l,a_l}_{v_q,a_q}\nn\\
 &-& {2e_q\over M^2} {1\over\sin^2\theta_W\cos^2\theta_W} 
     {\rm Re } \Big[D^{\tilde{Z}^\prime\tilde{Z}^\prime}(M^2)\Big]
     T_i|^{v'_l,a'_l}_{v'_q,a'_q}\nn\\
 &+& 2 {1\over\sin^4\theta_W\cos^4\theta_W} 
       {\rm Re } \Big[D^{\tilde{Z}^\prime\tilde{Z}^\prime}(M^2)
       {(M^2-m_Z^2) + i \Gamma_Z m_Z \over (M^2-m_Z^2)^2+(\Gamma_Zm_Z)^2} 
       \Big]
       T_i|^{v_l\,v'_l+a_l\,a'_l\,,v_l\,a'_l+v'_l\,a_l}
       _{v_qv'_q+a_qa'_q,v_qa'_q+va'_qa_q}~ \bigg\},~~~~~~
\label{eq:ME} \eea where $M$ is the invariant-mass of the produced lepton pair, $C_i$ is the
corresponding color factor and the functions $T_i|^{A_l,B_l}_{A_q,B_q}$ depend
on the kinematics of the process and on the order of the perturbative
calculation. We have defined the coefficients 
\bea
  A_f^{(\prime)} &=& a_f^{(\prime)2} + v_f^{(\prime)2},~ \nn\\
  B_f^{(\prime)} &=& 2 a_f^{(\prime)} v_f^{(\prime)},~
\eea and a modified propagator
$D^{\tilde{Z}^\prime\tilde{Z}^\prime}$ obtained by removing the photon and $Z$
contributions from the propagator $D^{\tilde{B}\tilde{B}}$ of Eq.\
(\ref{eq:prop}), since they have already been taken into account. The squared
$Z^\prime$-boson exchange as well as its interferences 
with the photon and $Z$-boson exchanges are included. We have adapted the MC@NLO
program of Ref.\ \cite{Fuks:2007gk} for the Drell-Yan process including $Z^\prime$
inspired by grand unified theories to the Heidi model case, which allow us to
match the matrix elements of Eq.\ (\ref{eq:ME}),  describing correctly the hard
parton emission by the initial state, to the HERWIG parton shower algorithm 
\cite{Corcella:2000bw} describing the soft and collinear emission. \\

In the considered process, the incoming quark, antiquark, and gluon can give rise
to an initial-state parton shower which is modeled in  HERWIG by starting with
the hard scattering partons and reconstructing backwards the preceding
branchings. For a specific splitting of partons $i\to jk$, the energy fraction
of parton $j$ is distributed according to the LO DGLAP splitting functions and
the phase space is restricted according to an angular ordered emission which
is based on the Sudakov form factor. This allows us to sum the
virtual corrections and unresolved real emissions to all orders and to correctly
treat the infrared singularities. The parton shower stops when a cutoff scale
defined by $(2.5 {\rm GeV})^2/E_i^2$ is reached, where a nonperturbative stage
is imposed. 

In order not to double count any contribution, the matching of the NLO matrix
element to the parton shower achieved in MC@NLO uses two separate
samples of Born-like and hard emission events which can have positive or
negative weight. They are statistically 
distributed according to the positive-definite corresponding part of the NLO
cross section and made separately finite by adding and subtracting the NLO part
of the expanded Sudakov form factor \cite{Frixione:2002ik}. The
total cross section is then given by their weighted sum averaged over the total
number of events.

\section{Analysis}\label{sec:3}


\subsection{Experimental constraints}\label{sec:3.1}

\begin{figure}
 \centering
 \includegraphics[width=0.32\columnwidth]{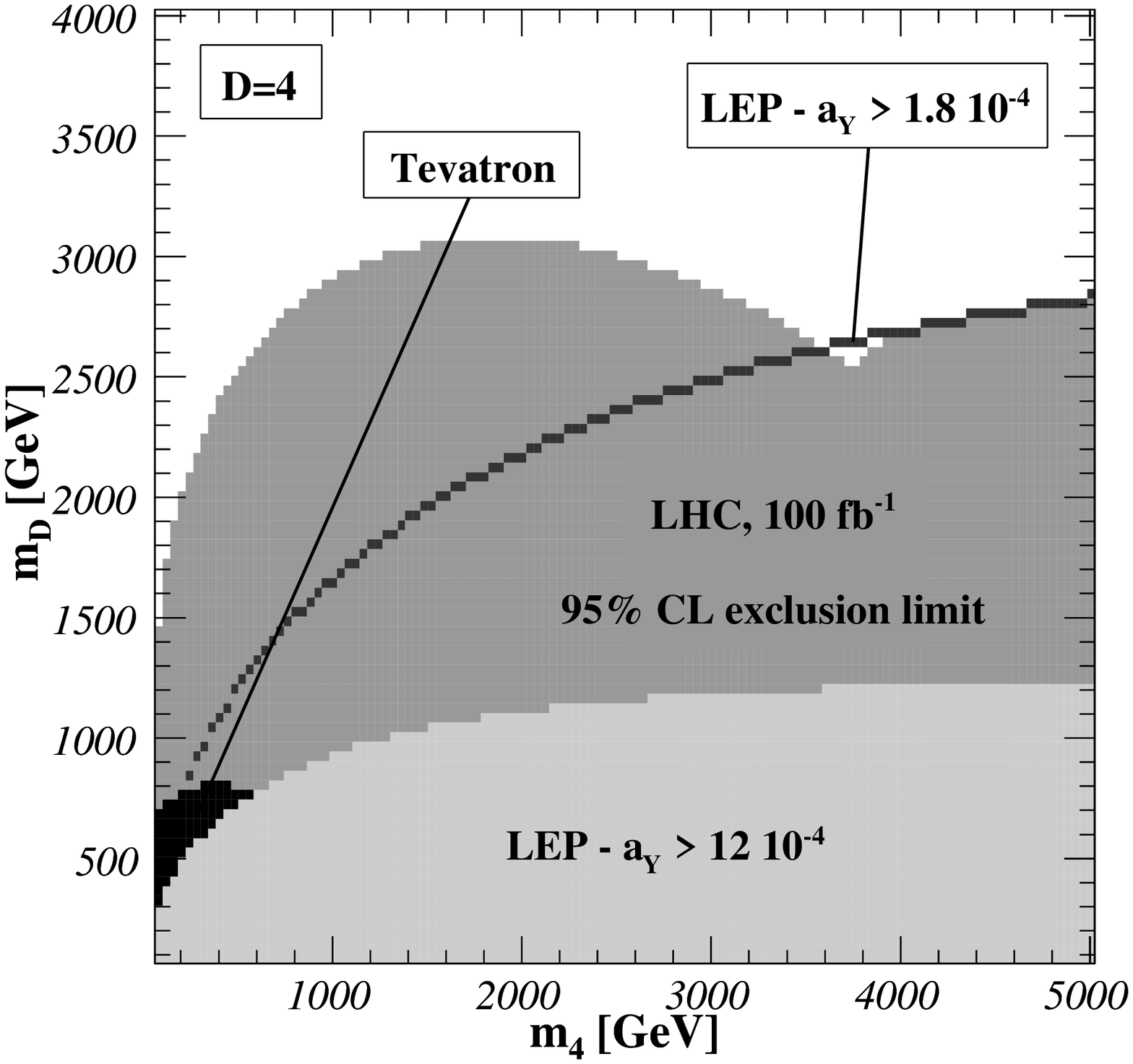}
 \includegraphics[width=0.32\columnwidth]{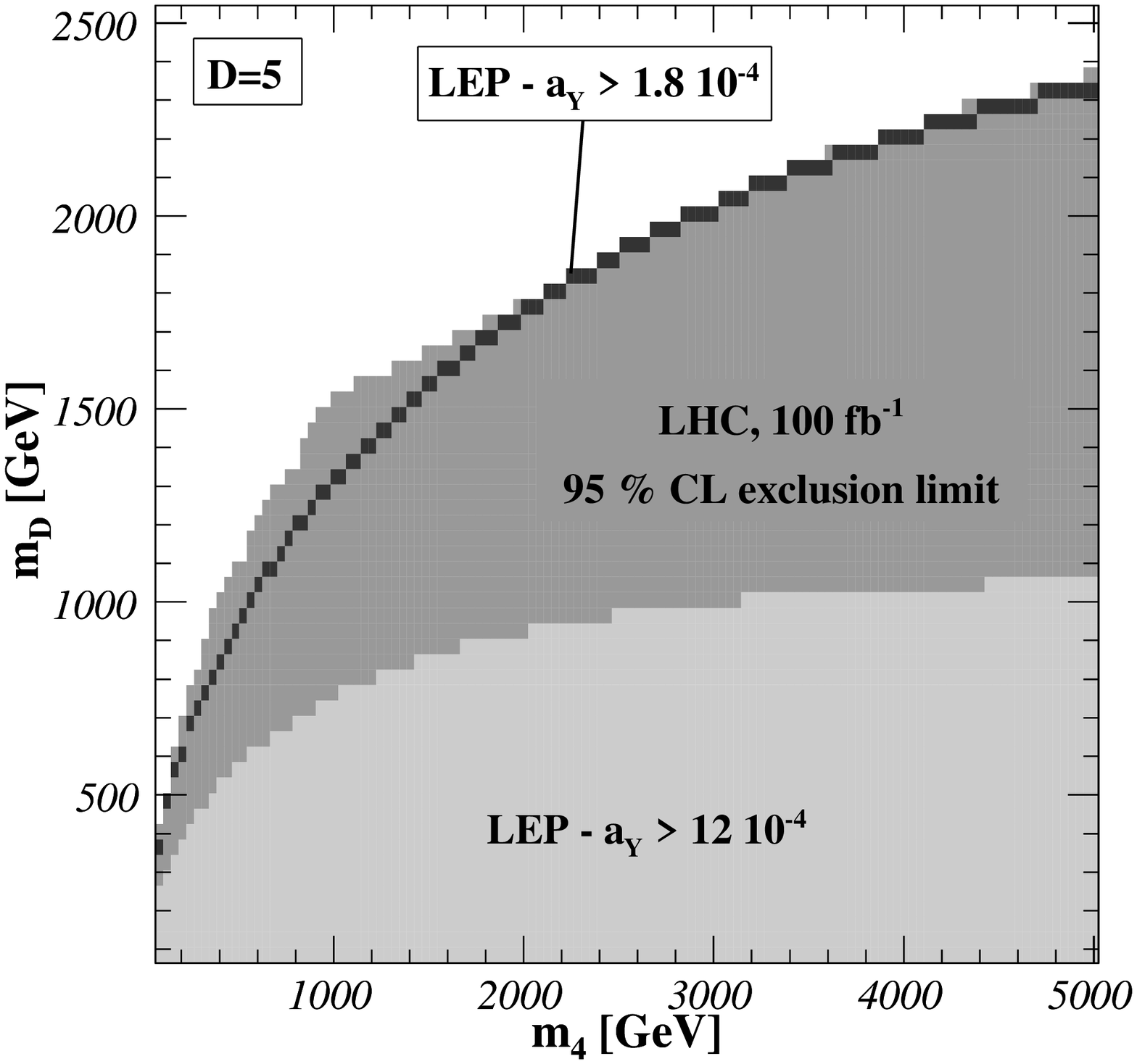}
 \includegraphics[width=0.32\columnwidth]{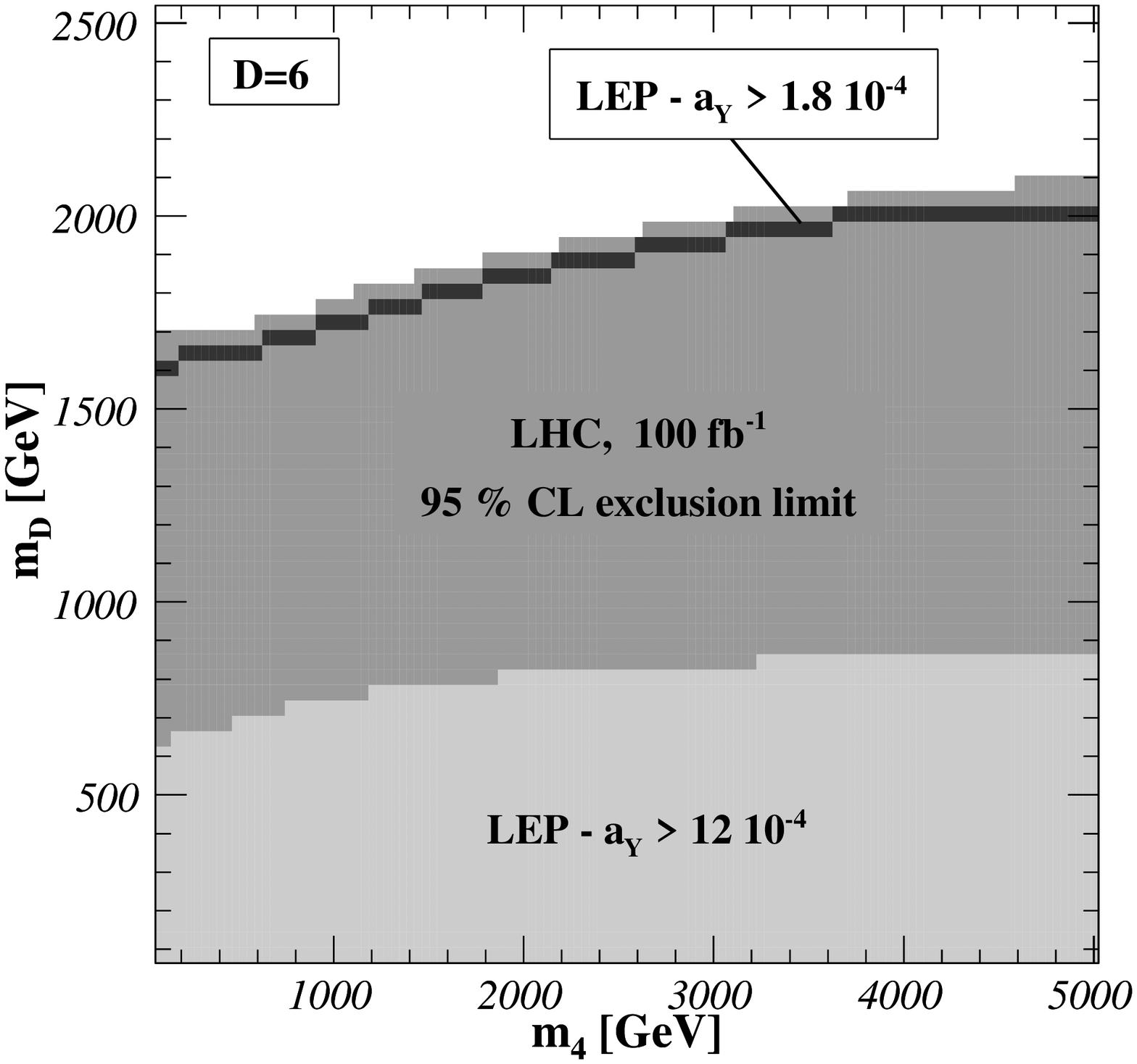}
 \caption{\label{fig:01}The $(m_4,m_D)$ plane for Heidi models with a
          four-dimensional (left), five-dimensional (center), and six-dimensional
	  (right) $Z^\prime$ boson. We show the regions excluded by the LEP
	  collider with (light gray region) and without (below the black line) leaving the
	  forward-backward asymmetry of the bottom quarks, by the Tevatron
	  (black region). We present also the region (dark gray) that the LHC
	  experiments will be able to exclude at the 95\% confidence level with
	  an integrated luminosity of 100 fb$^{-1}$. } 
\end{figure}

Although there is no experimental evidence for the existence of a 
 $Z^\prime$ boson, a number of data can be used to constrain the
parameter space. For example, limits on the parameter $a_Y$, and thus indirectly
on $m_4$ and $m_D$, can be obtained from precision measurements at the $Z$ pole,
strongly constraining the mixing between the standard model $Z$ boson and the new field.

Regarding the limits from the LEP experiment there
are two ways to proceed. One can take all the data
from the precision measurements \cite{lepewwg}
and try to make a fit. The problem here is that the
standard model is only barely compatible with these data
due to the inconsistency of the forward-backward asymmetry
$A_{FB}^b$ of the bottom quarks with the rest of the data.
Adding a Z' boson does not improve the situation, one
finds a limit $a_Y \leq 1.8\, 10^{-4}$ at the 95\% confidence level.
This problem with the data has led a number of authors
to consider fitting the data leaving out $A_{FB}^b$
\cite{Ferroglia:2004tt, Ferroglia:2002rg,Ferroglia:2004jx, Ferroglia:2006mj,
Chanowitz:2008ix, Chanowitz:2001bv, Chanowitz:2002cd}.
In this case a good fit to the data is possible 
if a Z' boson is present \cite{Ferroglia:2006mj,Chanowitz:2008ix}. 
With a Higgs mass ranging 
from $115$ to $495$ GeV, a larger range of $a_Y$
is allowed: $a_Y \leq 12\, 10^{-4}$. As the situation is somewhat controversial
we will present both limits in the following figures and discussions.

Imposing these limits on minimal Heidi models with three free
parameters, the number of dimensions 
$D$, the four-dimensional mass $m_4$ and the high-dimensional mass $m_D$,
 leads to the  excluded regions
shown in Fig.\ \ref{fig:01} for a four-dimensional (left), five-dimensional
(center), and six-dimensional (right) additional neutral gauge boson. The light
gray regions are the regions where $a_Y > 12\, 10^{-4}$, while the dark line
comes from the more stringent limits obtained by keeping all LEP data, i.e.\ it
corresponds to $a_Y=1.8\, 10^{-4}$. The excluded regions are thus those below
that line.

The best limits on the presence of a four-dimensional $Z^\prime$ boson come from the Tevatron collider. 
The principle guiding the search is straightforward.
One uses the decay of the $Z^\prime$ boson into an electron-positron pair
and looks for a peak in the invariant-mass. In addition one can use
information on the forward-backward asymmetry of the leptons. 
One therefore needs a prediction for the Drell-Yan production cross section.
 Moreover, when one is working within a specific
model, one can use the distribution of the leptons in the center-of-mass angle
$\cos{\theta^*}$ with respect to the beam axis, 
 leading to somewhat different limits for different models.
 This search has been made at the Tevatron \cite{cdf:2007sb} for the class 
of models discussed in Ref.\ \cite{Carena:2004xs}, but not for the specific models presented here.

The model dependence is  contained in the values of the vector
and axial-vector coupling constants. As argued in Ref. \cite{Carena:2004xs} 
it would be useful to have lower limits on the mass of the Z' boson presented
in the coupling constant plane. Unfortunately such a comparison has not been presented in
the literature. Instead we estimate the lower limits for the allowed mass 
of the $Z^\prime$-boson as a function $m_4$ and $m_D$, as follows. 
We use our implementation of the Heidi model in MC@NLO to predict
 the total cross section as a function of $m_4$ and $m_D$.
 Then, this cross section is normalized to the sequential $Z^\prime$ boson,
 which has a lower mass limit of $m_{Z^\prime}=923$ GeV \cite{cdf:2007sb}. 
We then connect the lower bounds on $m_{Z^\prime}$ within the 
 two studied models.
The results are given in the dark area of Fig.\ \ref{fig:01} (left panel). The derived
bounds 
 are of course somewhat qualitative,
 because the sensitivity to the $\cos{\theta^*}$ distribution 
is not exactly modeled this way. A precise analysis would require 
taking into account a bidimensional distribution, including the
 angular one as well, whereby one cannot ignore detector
effects \cite{cdf:2007sb}.
 However, this needs a detailed simulation of the detector 
and comparison with the actual data, which is beyond the scope of this paper.

For the higher-dimensional cases this procedure is not sufficient
and a bin-by-bin comparison of predicted and measured Drell-Yan cross sections is needed.
Unfortunately, these measurements are only presented in the literature \cite{Gelhaus:2005cd} for
a limited integrated luminosity. Within this limited statistics the Tevatron gives
no improvement over the LEP data. However, we note that in Ref. \cite{Abazov:2007ra}
there appears to be an excess in the $300-400$ GeV range, which might be consistent
with a spread-out Z' boson. We hope that published numbers
for the Drell-Yan cross section with a larger integrated luminosity
will appear soon, so that these
questions can be studied in more detail.

\renewcommand{\arraystretch}{1.4}
\begin{table}
 \caption{\label{tab:01} Heidi benchmark points accessible at the
 Tevatron and allowed by the LEP constraints after leaving out the
 forward-backward asymmetry of the bottom quarks.} 
 \begin{tabular}{c|ccc|}
 & $~~D~~$  & $~~m_4$ [GeV]$~~$ & $~~m_D$ [GeV]$~~$\\
 \hline
 A & 5 & 100 & 300 \\
 B & 6 & 50  & 650 \\
 \end{tabular}
\end{table}

\begin{figure}
 \centering
 \includegraphics[width=0.49\columnwidth]{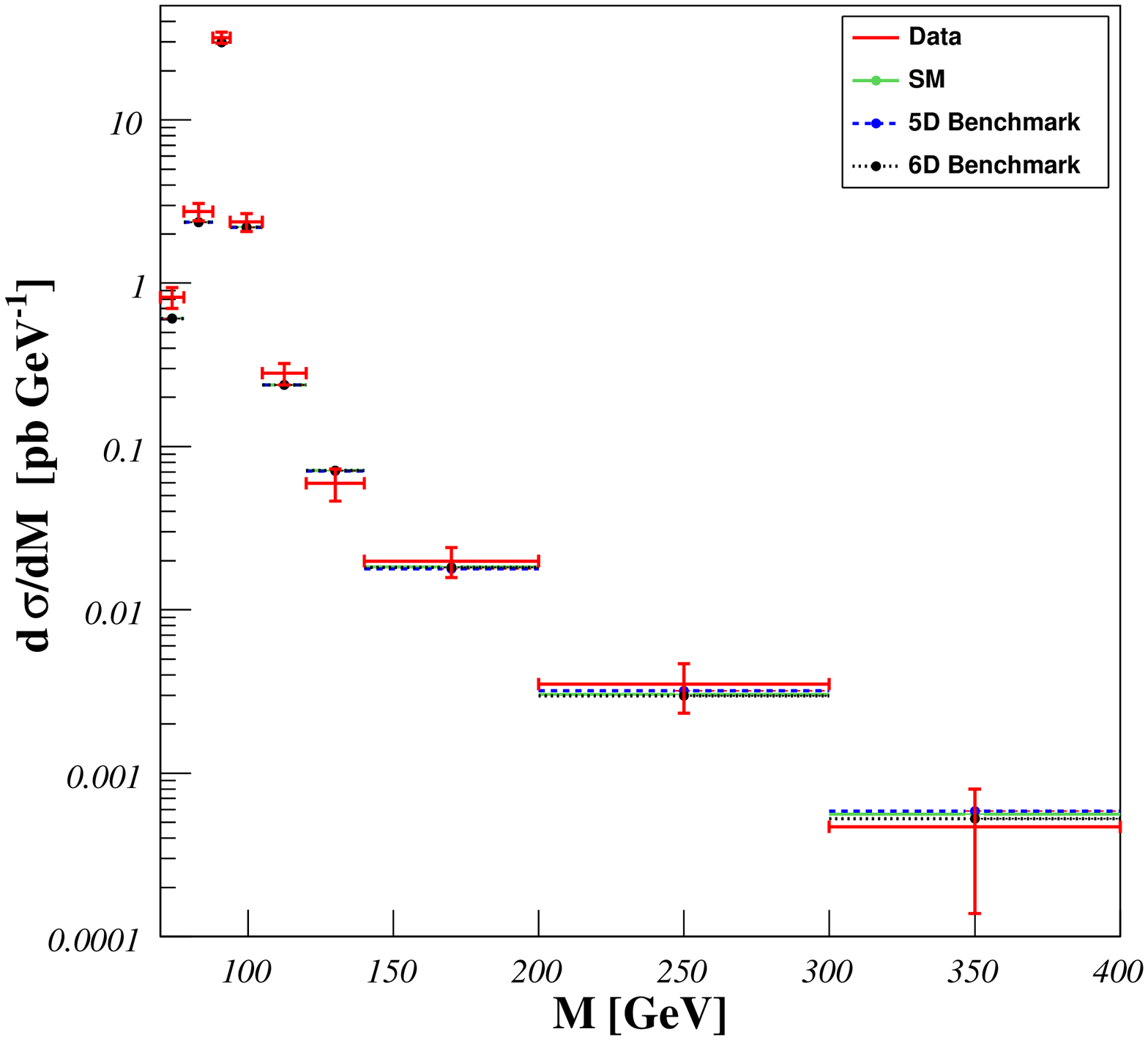}
 \includegraphics[width=0.49\columnwidth]{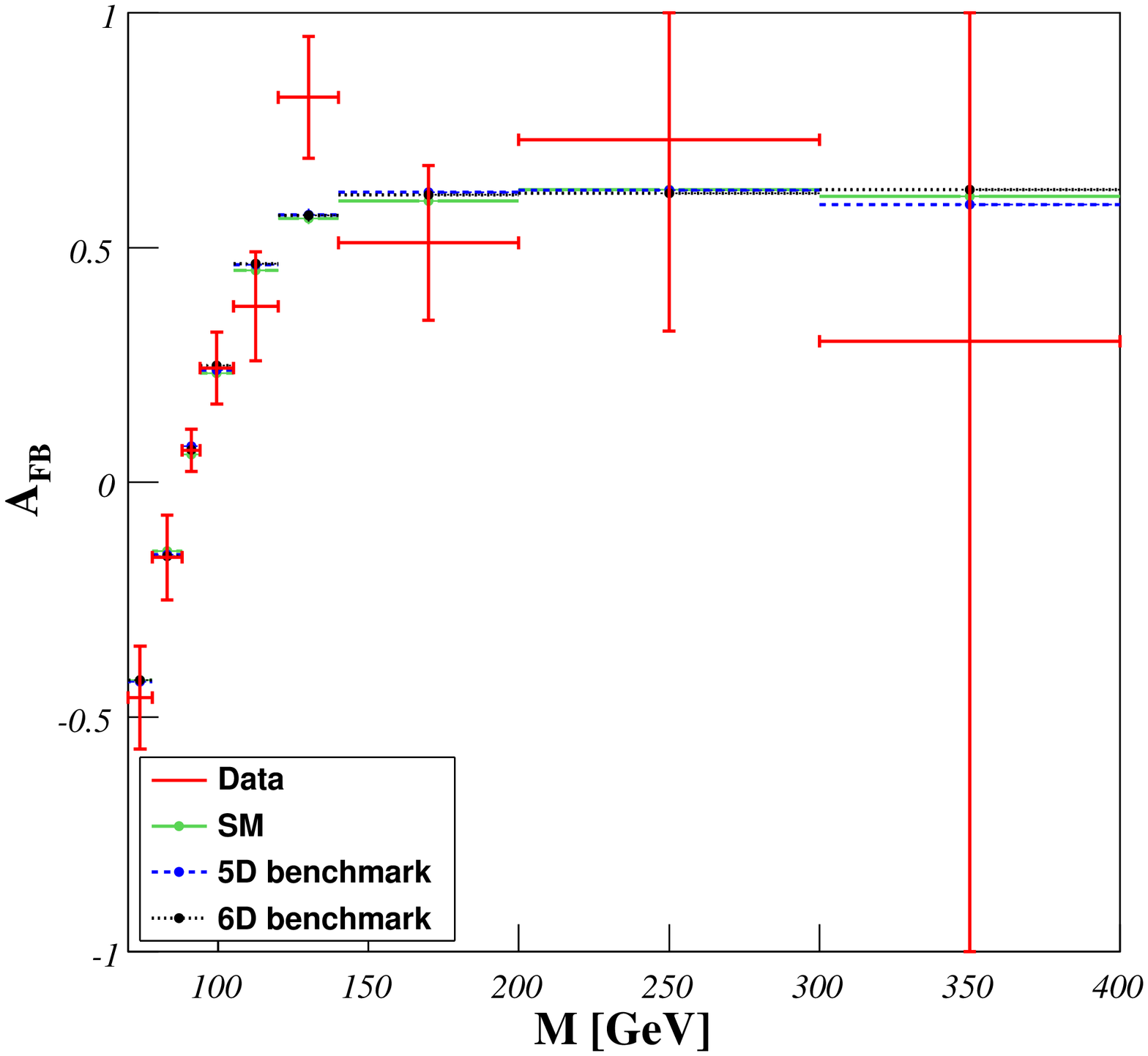}
 \caption{\label{fig:02}Drell-Yan differential cross section and
   forward-backward asymmetry with and without Heidi $Z^\prime$ contributions at
   the Tevatron. The results are presented for the two benchmark points
    of Table \ref{tab:01}.}
\end{figure}

In Fig.\ \ref{fig:02}, we compare our results  with the limited statistics data 
published in Ref.\ \cite{Gelhaus:2005cd}. We show both  the
differential cross section $\d \sigma / \d M$ and the forward-backward asymmetry. As an
example, we chose the two Heidi benchmark points presented in Table \ref{tab:01},
which are just within the region of the parameter space allowed by the constraints
coming from LEP, provided that the forward-backward asymmetry of the bottom quarks
is left out, and  ``friendly'' for the Tevatron. We show our predictions for
the Heidi model together with those of 
the standard model, i.e.\ when no additional
$Z^\prime$ boson is present. We use the CTEQ6M (NLO $\overline{\rm MS}$)
\cite{Pumplin:2002vw} sets for the parton densities in the proton
and antiproton. As can be seen from the figures, all the predictions are
compatible with the data for both observables. However, data with a higher
luminosity would make it sensible to divide the studied invariant-mass range in smaller
bins, which could be used to emphasize deviations from the standard model.

\subsection{Predictions for the LHC}\label{sec:3.2}

In this section, we discuss the sensitivity of the LHC experiments to Heidi
models, considering four-dimensional, five-, and six-dimensional
extra neutral gauge bosons. The discovery potential of the LHC is determined
through the investigation of possible deviations from 
the standard model predictions in the tail of the Drell-Yan invariant-mass
distribution, the studied range being 
\bea \label{eq:range}
200 {\rm~GeV} \leq M \leq  4500 {\rm~GeV}~.~
\eea

The main source of background for Drell-Yan $l^+l^-$ events is QCD
multijet and direct photon production where the jets have a large 
electromagnetic component. However, with typical experimental selection cuts one can
obtain a signal almost free from any QCD background, its part to the
measurements being reduced to at most 1\% 
\cite{atlas:1999fq,atlas:1999fr, Ball:2007zza}. For that reason, we apply the
following cuts
\bea \label{eq:cuts}
  \left\{ 
    \begin{array}{l c l} 
      p_T &>& 20 {~\rm GeV}\\
      \big|\eta\big| &<& 2.5
    \end{array}~,~
  \right.
\eea to both leptons, which allows us to consider as a new physics signal any
significant excess of $l^+l^-$ events with respect to the standard model Drell-Yan
irreducible background, neglecting thus any other source of background.

\renewcommand{\arraystretch}{1.4}
\begin{table}
 \caption{\label{tab:02}Typical Heidi benchmark points accessible at the
 LHC and allowed both by the Tevatron and by the LEP constraints, after leaving out
 the forward-backward asymmetry of the bottom quarks.} 
 \begin{tabular}{c|ccc|}
 & $~~D~~$  & $~~m_4$ [GeV]$~~$ & $~~m_D$ [GeV]$~~$\\
 \hline
 C & 4 & 500 & 300 \\
 D & 5 & 200 & 400 \\
 E & 6 & 50  & 700 \\
 \end{tabular}
\end{table}

For each bin $i$ of the investigated range of Eq.\ (\ref{eq:range}), we calculate the 
quantities $N_i^\prime$ and $N_i$, representing the number of events related to
Drell-Yan-like lepton-pair production with and without an additional neutral
gauge boson, respectively. For a given 
luminosity $L$, they can be derived from the differential cross section by
\bea
  N_i^{(\prime)} = \int_{M_1}^{M_2} \d M \frac{\d \sigma^{(\prime)}}{\d M} L,~
\eea  
where $M_1$ ($M_2$) is the lower (upper) limit of the considered bin, and $\d
\sigma$ ($\d \sigma^\prime)$ the Drell-Yan cross section without (with) an
additional $Z^\prime$ gauge boson. Expecting the standard model Drell-Yan
background, we can
estimate the significance of a possible Heidi signal through the quantity 
\bea
  \chi^2 = \sum_{i=1}^{n} \frac{\big(N_i-N_i^\prime\big)^2}{N_i}
\eea 
where $n$ is the number of bins which the range of Eq.\ (\ref{eq:range}) is
divided in. The LHC will then be able to exclude a given point of the Heidi parameter
space at the 95\% confidence level if the corresponding $\chi^2$ is bigger than
21. \\

In order to take advantage of the differences between the standard model and the
Heidi predicted line shape of the Drell-Yan mass spectrum in the high
invariant-mass regions, we divide the studied range into nine bins of 100 GeV for
200 GeV $\leq M \leq $ 
1100 GeV and we take three larger bins at very high invariant-mass in order to
have higher statistics, i.e.\ 1100 GeV $\leq M \leq 1300$ GeV, 1300 GeV $\leq M
\leq 1600$ GeV, and 1600 GeV $\leq M \leq 4500$ GeV. That allows us to avoid
spurious sensitivities since we have then at least 50 expected events for each
bin.
This way we have a good idea of the typical reach of the LHC in the parameter space of the
models. The results shown in Fig.\ \ref{fig:01} use cross sections calculated at a
center-of-mass energy of 14 TeV with again the CTEQ6M
(NLO $\overline{\rm MS}$) \cite{Pumplin:2002vw} sets of parton densities, assuming
a LHC lumisotiy of 100 fb$^{-1}$.\\

For the four-dimensional model, the LHC will be able to reach a mass range of
several TeV, as can be seen from the dark gray region in the left panel of Fig.\
\ref{fig:01}. This range 
can be divided in two areas, one at small value of $m_4$ and one at larger values.
The first zone corresponds to a region where the $Z^\prime$ mass is smaller than or
equal to about 3 TeV and is fairly typical since it is also found for other models
with electroweak size couplings. In contrast, the LHC-reachable zone where $m_4$ is larger
corresponds to a heavier $Z^\prime$ but with a much larger width due to the large
value of the mixing parameter $a_Y$, which
makes it possible to detect the new vector boson via its interferences with the standard model
photon and Z boson. In comparison to the results of previous
experiments, the LHC will allow us to considerably enlarge
the model parameter space that can be investigated in the four-dimensional case.
For the
five-, and six-dimensional cases, however,  the LHC will not be able
to cover a parameter space region much larger than is already constrained
by  the LEP collider. This is the case if we take the strict limits that exist if we keep the
data coming from the forward-backward asymmetry of the bottom quarks in the
electroweak precision fit. As stated before, the
resonance is  replaced by a massive continuum, rendering the detection of any
excess in the Drell-Yan distributions more complicated and restricting the
regions of  the parameter space that can be reached at the LHC, as can be seen in
the central and right panels of Fig.\ \ref{fig:01}. However, if one prefers to take
the more conservative
limits obtained by leaving out the forward-backward asymmetry of the bottom quarks,
the LHC is sensitive to a range of parameters roughly double  
the size  that is reached by previous colliders.\\

\begin{figure}
 \centering
 \includegraphics[width=0.32\columnwidth]{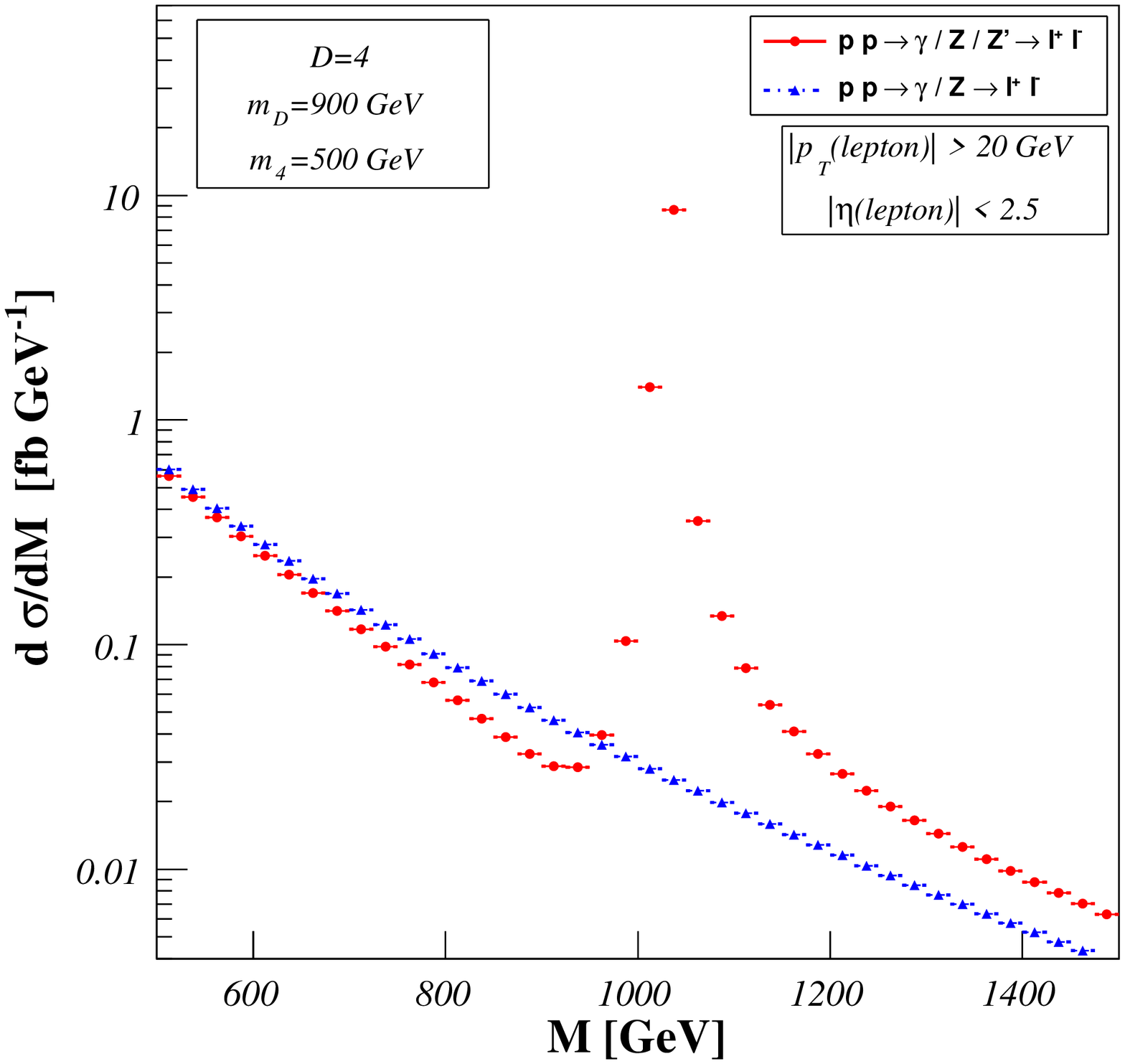}
 \includegraphics[width=0.32\columnwidth]{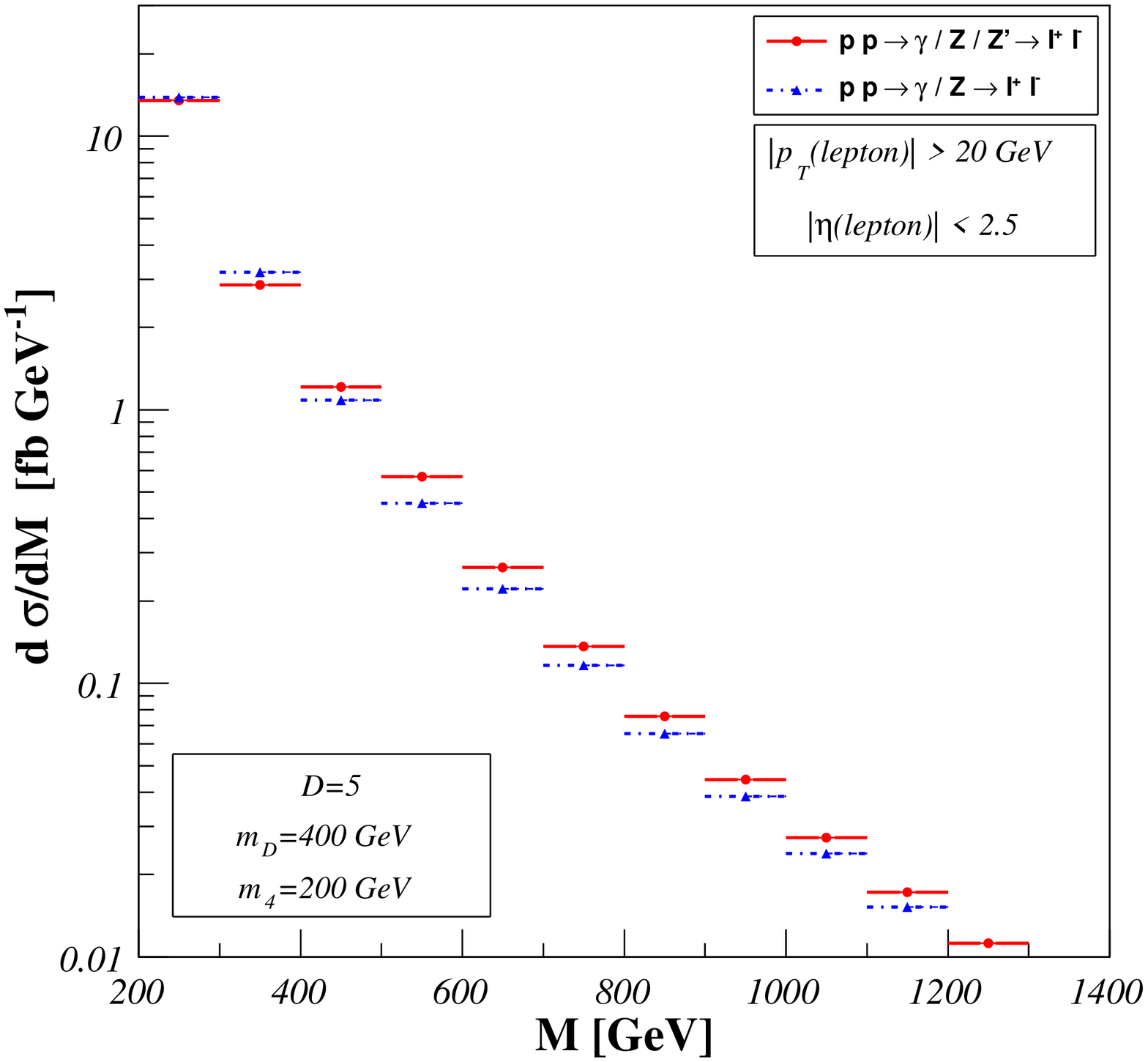}
 \includegraphics[width=0.32\columnwidth]{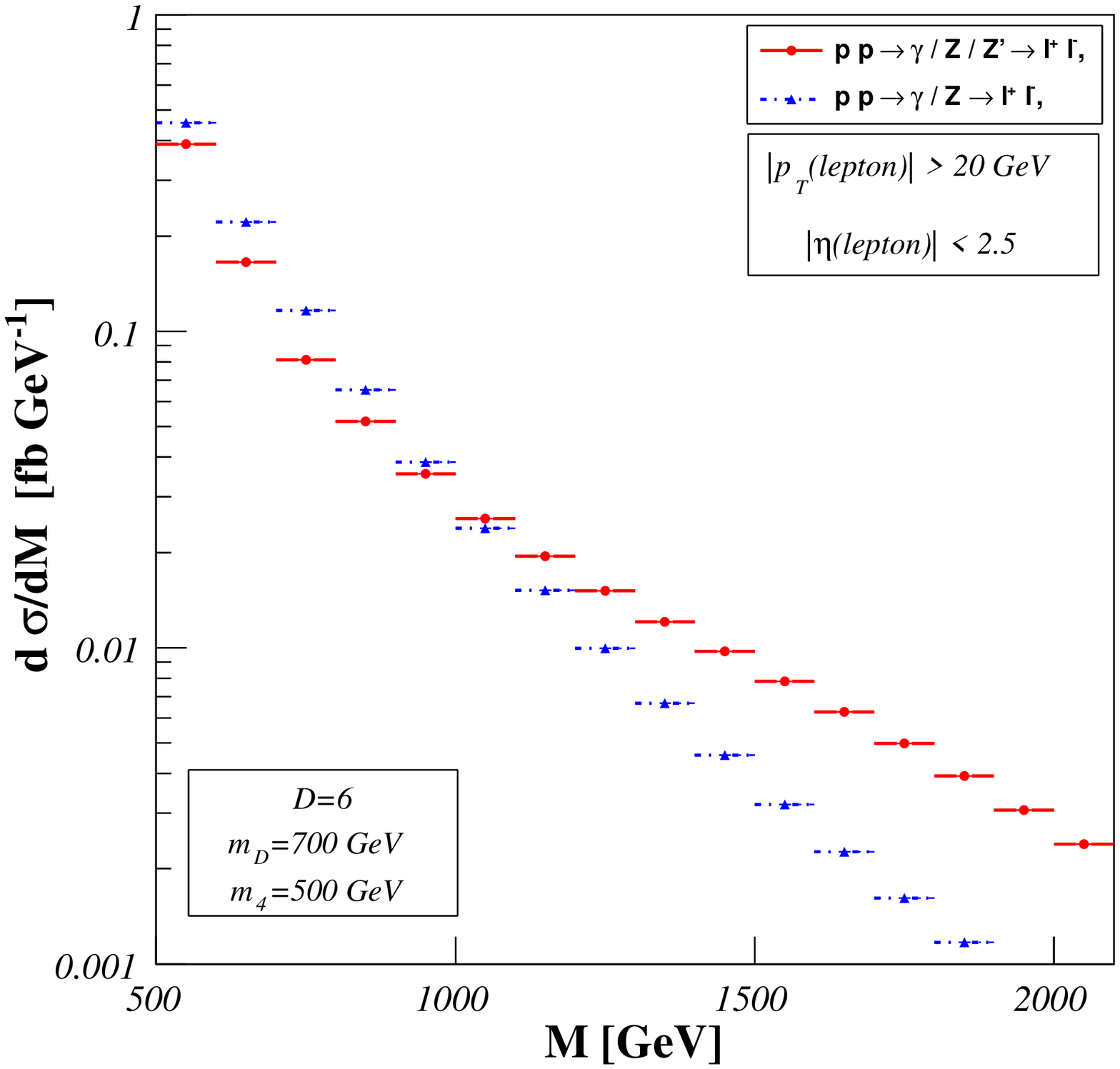}
 \caption{\label{fig:03} Drell-Yan differential cross sections with (solid red) and
 without (dashed blue) additional Heidi $Z^\prime$ bosons at the LHC for the
 benchmark points presented in Table\ \ref{tab:02}. The 
 typical experimental cuts of Eq.\ (\ref{eq:cuts}) are applied.
 } 
\end{figure}

As an example, we show in Fig.\ 
\ref{fig:03} lepton-pair invariant-mass spectra for the benchmark points presented
in Table \ref{tab:02}. We compare
MC@NLO predictions including a Heidi $Z^\prime$ boson (solid red curves) with
the ones expected in the standard model case (dashed blue curves). 
In the
four-dimensional model (left panel of Fig.\ \ref{fig:03}), far from the peak region located at the $Z^\prime$ mass
given by Eq.\ 
(\ref{eq:zpmass}), the spectra with and without $Z^\prime$ will of course
coincide, while around the $Z'$ mass, we have a clear resonance due to the presence
of the extra field. For five-dimensional (central panel of Fig.\ \ref{fig:03}) and
six-dimensional (right panel of Fig.\ \ref{fig:03}) extra neutral gauge 
bosons, the
resonance is replaced by a massive continuum, which makes the detection a bit more
tricky. However, large deviations can still easily be detected if we consider the
line shapes of the distributions rather than the total number of events.

\section{Conclusions}
\label{sec:4}

We studied a new class of renormalizable models containing 
a $Z'$ boson with a nontrivial
K\"all\'en-Lehmann representation, derived from the mixing with an inert
higher-dimensional field. Existing LEP limits constrain this class of models,
but the amount depends on the interpretation of the analysis of the
precision data, allowing for strict or  loose limits. In the case of the loose
limits we found, using a state of the art MC@NLO program, that the LHC 
will be able to study a large range of the allowed parameter space.   
In the case of the strict limits even the LHC will have difficulty to
improve on existing bounds, but there  may be some hope for a large luminosity 
option. Present bounds from the Tevatron, which may be relevant at 
low masses, are not satisfactory, since only a small fraction of the data
have been analyzed for the Drell-Yan cross section. We hope the situation
will improve soon.

\acknowledgments 
The authors acknowledge F. Chevallier, B. Cl\'ement and Q. Li for
useful discussions.


\end{document}